\documentclass[fleqn,twoside]{article}
\usepackage{espcrc2}

\usepackage{graphicx}

\title{Probing nonstandard interactions with reactor
  neutrinos\thanks{The talk was presented by T.~I.~Rashba at the
    Neutrino Oscillation Workshop (NOW 2008) (Conca Specchiulla,
    Otranto, Lecce, Italy), Sept. 6-13, 2008.  T.I.R. is grateful to
    the organizers of NOW 2008 for the invitation and fruitful
    atmosphere of the Conference.  Authors thank M.~Deniz and H.~Wong
    for valuable and productive discussions. This work has been
    supported by CONACyT (Mexico), SNI-Mexico, PAPIIT (Mexico) grant
    IN104208, DFG (Germany) grant SFB/TR 27, Russian foundation for
    basic research and RAS Program ``Solar activity''.  T.I.R. thanks
    also Physics Department of CINVESTAV for the hospitality during
    the visit when part of this work was done.}}

\author{J. Barranco\address{Max-Planck-Institut f\"ur Gravitationsphysik
  (Albert-Einstein-Institut), Am M\"uhlenberg 1, D-14476 Golm,
  Germany},
O. G. Miranda\address{Departamento de F\'{\i}sica, Centro de
  Investigaci{\'o}n y de Estudios Avanzados del IPN, Apdo. Postal
  14-740 07000 M\'exico, D F, M\'exico}
and 
T. I. Rashba\address{Max-Planck-Institut f\"ur Physik
  (Werner-Heisenberg-Institut), F\"ohringer Ring 6, D-80805 M\"unchen,
  Germany}~\address{Pushkov Institute of Terrestrial Magnetism,
  Ionosphere and Radio Wave Propagation of the Russian Academy of
  Sciences (IZMIRAN), 142190, Troitsk, Moscow region, Russia}}

\begin{document}

\begin{abstract}
New limits on the weak mixing angle and on the electron neutrino
effective charge radius in the low energy regime, below 100~MeV, are
obtained from a combined fit of all electron-(anti)neutrino electron
elastic scattering measurements. We have included the recent TEXONO
measurement with a CsI (Tl) detector. 
Only statistical error of this measurement has been taken into
account.
Weak mixing angle is found to be $\sin^2\theta_W=0.255_{-0.023}^{+0.022}$.
The electron neutrino effective charge radius squared is bounded to be
$\langle r_{\nu_e}^2 \rangle = (0.9_{-1.0}^{+0.9})\times10^{-32}$~cm$^{2}$.
The sensitivity of future low energy neutrino experiments to
nonstandard interactions of neutrinos with quarks is also discussed.
\end{abstract}

\maketitle

\section{Introduction}

Current and future measurements of elastic neutrino scattering off
electrons and quarks in reactor and accelerator experiments are
getting more and more precise. This opens a possibility to search for
possible nonstandard neutrino interactions (NSI) in such experiments.

Nonstandard neutrino interactions  with quarks and electrons are still
allowed to be rather large. It was shown recently that coherent neutrino
nuclei scattering is a very sensitive probe to
NSI~\cite{Barranco:2005yy,Barranco:2007tz}.  The sensitivity to some specific
NSI models, such as extra heavy neutral gauge bosons, leptoquarks,
supersymmetry with broken R-parity, could be better than present constraints
and, therefore, could give complementary bounds to future LHC
limits~\cite{Barranco:2007tz}.

Here we focus on the NSI effects in neutrino detection at low
energies~\cite{Barranco:2005yy,Barranco:2007tz,Barranco:2007ej,Barranco:2007ea,Barranco:2008tk}.
We will present the limits on the weak mixing angle, $\sin^2\theta_W$,
and on the electron neutrino effective charge radius, $\langle
r_{\nu_e}^2 \rangle$, obtained from the combined analysis of all
available low energy (anti)neutrino-electron elastic scattering
experiments including recent $\nu$-$e$ cross section measurement by
TEXONO~\cite{Deniz:2008rw}.

\section{Weak mixing angle and effective electron neutrino charge radius}

The straight-forward definition of a neutrino charge radius in the
Standard Model has been proved to be gauge-dependent and, hence,
unphysical. There have been recent discussions in the literature on
the possibility to define and to calculate a physically observable
neutrino charge radius~\cite{Bernabeu:2002nw,Fujikawa:2003tz}. All
relevant references can be found in Ref.~\cite{Barranco:2007ea}.  A
more general interpretation of the experimental results which we adopt
here is that they are limits on certain nonstandard contributions to
neutrino scattering.

All available data of the (anti)neutrino-electron scattering from the
following reactor and accelerator experiments were analyzed: first
measurement of neutrino-electron scattering made by Reines, Gurr and
Sobel (Irvine), the Kurchatov institute group at the Krasnoyarsk
reactor, the group from Gatchina at the Rovno reactor, MUNU at the
Bugey reactor, recent TEXONO measurement~\cite{Deniz:2008rw}, LAMPF and LSND.
%
%
All references to the experimental data used in the analysis are given
in Ref.~\cite{Barranco:2007ea}, except the recent talk by the TEXONO
Collaboration at the ICHEP 2008 Conference~\cite{Deniz:2008rw}. The
measurement presented by TEXONO Collaboration was obtained using
CsI(Tl) detector at Kuo-Sheng Nuclear Power Plant in
Taiwan~\cite{Deniz:2008rw}
\begin{equation}
\sin^2\theta_W=0.24\pm0.05~\mbox{(stat)}\,.
\end{equation}
We would like to note that only statistical error was presented by
now, therefore our new global limits should be considered as the
preliminary ones.

\begin{figure}
\begin{center}
\includegraphics[width=\columnwidth]{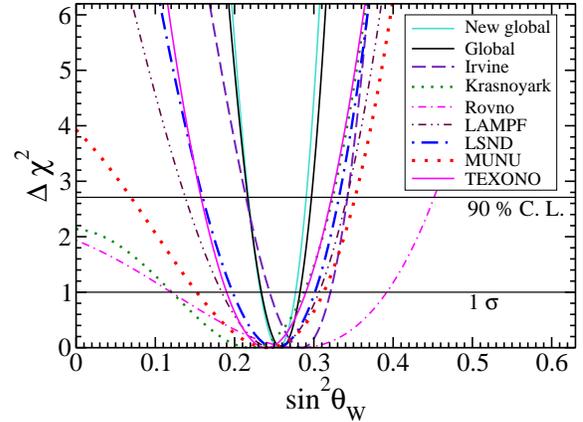}
\end{center}
\vskip-1.2cm
\caption{\label{fig:sin.eps}$\chi^2$ fits of the weak mixing angle for various
  experiments are show. ``New global'' means with TEXONO 2008, while
  ``global'' means without TEXONO 2008.}
\end{figure}

\begin{figure}
\begin{center}
\includegraphics[angle=90,width=\columnwidth,height=0.33\textheight]{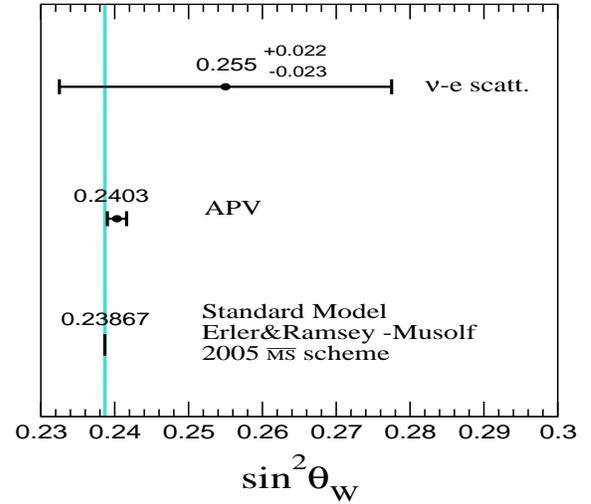}
\end{center}
\vskip-1.2cm
\caption{\label{fig:sindiag3.eps}The weak mixing angle in low energy
  regime below 100~MeV is shown. For comparison the atomic parity
  violation (APV) measurement~\cite{Bennett:1999pd} and the Standard
  Model prediction~\cite{Erler:2004in} are also presented.}
\end{figure}

The data of (anti)neutrino-electron scattering experiments listed
above were used (see Figure~\ref{fig:sin.eps} and \ref{fig:sindiag3.eps})
to obtain a new limit on the weak mixing angle with
better than 10\% precision at energies below
100~MeV
\begin{equation}
\label{eq:expweak}
\sin^2\theta_W=0.255_{-0.023}^{+0.022}\,.
\end{equation}
This limit slightly improves our previously published result obtained
in Ref.~\cite{Barranco:2007ea} due to the inclusion of recent TEXONO
2008 measurement with CsI(Tl) detector~\cite{Deniz:2008rw}.
It is not competitive with the current best measurements at low
energies of the atomic parity violation~\cite{Bennett:1999pd} and of
the M{\o}ller scattering by SLAC E158~\cite{Anthony:2005pm}, both
experiments have a precision better than 0.5\%.  However, it is derived
from a different channel and it could therefore give new information
about other effects, such as the electron neutrino effective charge
radius.

This value of the weak mixing angle, Eq.~(\ref{eq:expweak}), and the
Standard Model prediction~\cite{Erler:2004in}
\begin{equation}
\sin^2\theta_W=0.23867\pm0.00016
\end{equation}
were used to set a new limit to the electron neutrino effective charge
radius squared
\begin{equation}
\langle r_{\nu_e}^2 \rangle = (0.9_{-1.0}^{+0.9})\times10^{-32}$~cm$^{2}\,.
\end{equation}
This result improves previous
bounds~\cite{Yao:2006px,Barranco:2007ea}.
We remind that only statistical error for the TEXONO
measurement~\cite{Deniz:2008rw} was taken into account.

One can conclude, that from the measurement of the weak mixing angle
with 1\% precision it is possible to find a strong evidence for the
electron neutrino charge radius, which is estimated theoretically to
be of the order of $0.4\times10^{-32}$~cm$^2$~\cite{Bernabeu:2002nw}.

\section{Neutrino-quark NSI}

We have demonstrated that neutrino coherent scattering off nuclei has
a lot of potential to make precision tests of new physics, in
particular, to probe NSI~\cite{Barranco:2005yy,Barranco:2007tz} that
naturally appear in many extensions of the Standard Model.
A new experiment would be able to test large NSI parameters and
therefore may exclude new solutions to the solar neutrino
data~\cite{Miranda:2004nb}.

As concrete proposed experiments, we have
discussed~\cite{Barranco:2007tz}: TEXONO~\cite{Wong:2008vk}, the
stopped pion source with a noble gas detector~\cite{Scholberg:2005qs}
and the beta beams~\cite{Bueno:2006yq}. We would like to note that
currently there are other experimental
proposals~\cite{Barbeau:2007qi}, which are not covered by our
analysis.

\section{Summary}

The weak mixing angle with precision better than 10\% at energies
below 100~MeV was obtained using all (anti)neutrino electron
scattering measurements. To get this result the combined set of all
available data from accelerator (LSND and LAMPF) and reactor (Irvine,
Rovno, Krasnoyarsk, MUNU and TEXONO) experiments was used. This
analysis was also applied to set a new limit to the electron neutrino
effective charge radius squared which improves previous bounds.

We have discussed that neutrino coherent scattering off nuclei has a
lot of potential to make precision tests of new physics, in
particular, to test nonstandard neutrino-quark interactions.

\end{document}